\documentclass[aps,nofootinbib,twocolumn,showpacs,pra,10pt]{revtex4-1}
\usepackage{amsfonts}
\usepackage{amsmath}
\usepackage{amssymb}
\usepackage{graphicx}

\DeclareMathOperator{\Trace}{Tr}

\usepackage{color}

\newcommand{\ra}{R_{a}}
\newcommand{\rd}{R_{d}}
\newcommand{\rt}{R_{t}}

\newcommand{\ea}{E_{a}}
\newcommand{\ed}{E_{d}}
\newcommand{\et}{E_{t}}
\newcommand{\zphi}{Z_d}
\newcommand{\zchi}{Z_t}
\newcommand{\tatom}[4]{ T_{ #1, #2, #3, #4 }^{a} }
\newcommand{\tdimer}[4]{ T_{ #1, #2, #3, #4 }^{d} }
\newcommand{\ttrimer}[4]{ T_{ #1, #2, #3, #4 }^{t} }
\newcommand{\tgen}[5]{ T_{ #1, #2, #3, #4 }^{ #5 } }

\begin{document}

\title{Universal behaviour of four-boson systems from a functional 
renormalisation group}
\author{Benjam\'in Jaramillo \'Avila and Michael C. Birse}
\affiliation{Theoretical Physics Division, School of Physics and Astronomy,\\
The University of Manchester, Manchester, M13 9PL, UK\\}

\begin{abstract}
We apply a functional renormalisation group to systems of four bosonic atoms 
close to the unitary limit. We work with a local effective action that includes
a dynamical trimer field and we use this field to eliminate structures that do 
not correspond to the Faddeev-Yakubovsky equations.
In the physical limit, we find three four-body bound states below the shallowest 
three-body state. The values of the scattering lengths at which two of these 
states become bound are in good agreement with exact solutions of the four-body
equations and experimental observations. The third state is extremely shallow.
During the evolution we find an infinite number of four-body states based 
on each three-body state which follow a double-exponential pattern in the 
running scale. None of the four-body states shows any evidence of dependence on 
a four-body parameter.
\end{abstract}

\pacs{67.85.-d, 03.65.Ge, 11.10.Hi, 21.45.-v}
\maketitle

\vskip 10pt

\section{\label{section:introduction}Introduction}
Systems where two-body scattering lengths are much longer than ranges of the 
forces between the particles are important in various areas of physics.
Their low-energy properties display universal scaling behaviour, controlled
by the ``unitary limit" in which the scattering length tends to infinity.
In nuclear physics, the large scattering lengths are large enough that 
low-energy aspects of few-nucleon systems can be described in this framework
\cite{bvk02,hp10}. In atomic physics, the shallow dimer of $^4$He atoms leads to 
a scattering length that is about 100 times larger than the size of the atoms 
\cite{gsthks00}. Even better examples are provided by ultra-cold atoms in traps, 
where Feshbach resonances can be used to tune the scattering lengths to values 
very close to the unitary limit \cite{cgjt10}. 

In the unitary limit, three-boson systems display a remarkable effect, first 
predicted by Efimov in 1970 \cite{efim70,efim71}. They possess an infinite tower 
of three-body bound states, with energies in a constant ratio of $\sim 515.0$. 
This breaks the expected scale invariance to a discrete symmetry, with one 
three-body parameter
needed to fix the energies of all these states. In real systems, the sequence of
deeply bound states is cut off by the range of the forces, and the shallowest ones 
by the finite scattering length. Three-fermion systems can also show Efimov 
behaviour, provided there are enough species to allow spatially symmetric states. Although there were suggestions that the $A=3$ nuclei $^3$H and $^3$He could 
be interpreted as Efimov states \cite{bhvk00}, the first clear observation of
such states was in an ultra-cold gas of caesium atoms \cite{k06}. Reviews 
of the field can be found in Refs.~\cite{bh06,fzbhng11,hp11}.

This behaviour in the three-body sector feeds through to four-body systems, 
where most numerical calculations find two bound states in each Efimov cycle 
\cite{hp07,vsdg09,delt10} whose energies are fixed ratios to the nearest
three-body state. However, in contrast, Hadizadeh \textit{et al.}~find 
up to three four-body states per cycle, with energies that depend on an 
additional four-body parameter \cite{had11,had12}, supporting their 
earlier results of Ref.~\cite{ytdf06}. Experimental evidence for two 
four-body states based on an Efimov three-body state has been seen in the 
recombination rates of trapped $^{133}$Cs atoms \cite{f09}, with resonances 
that are consistent with the results of Refs.~\cite{hp07,vsdg09,delt10}.

Renormalisation-group methods have been applied to elucidate scaling 
behaviour in few-body systems \cite{bhvk99,bhvk00,bb05,nish08,hp11} and hence to
determine their relevant parameters. Here we apply a functional renormalisation 
group (FRG) \cite{wett93,btw02} to the four-boson problem. During the evolution 
we observe a double-exponential pattern of four-body states built on each 
three-body state, similar to the ``super-Efimov" behaviour found by Nishida, Moroz 
and Son in a two-dimensional three-body system \cite{mns13}. These have energies 
that can be expressed in terms of a universal scaling function, similar to that 
in Refs.~\cite{had11,had12}, but they show no evidence of dependence on an 
additional four-body parameter. The states in our ``super-Efimov" pattern are 
not necessarily physical and, away from the unitary limit, we find that only three 
of them are present in the last Efimov cycle and so can appear as physical bound 
states. The two deepest of this states appear for scattering lengths that are 
in reasonable agreement with those found in studies of four-body equations 
\cite{vsdg09,delt10} and experimental observations \cite{f09}.

This paper is structured as follows.
In Section \ref{section:FRG_and_running_action} we present the FRG and 
running action that we use to study four-atom systems. Previous results
on the three-body sector are summarised in Sec.~\ref{section:three_body},
as they provide key input into our four-body equations. Those equations
are presented in Sec.~\ref{section:four_body} together with our results for 
the four-body sector. We summarise and conclude in 
Sec.~\ref{section:conclusions}.

\section{\label{section:FRG_and_running_action}FRG and running action}

The FRG we use is based on a running version of effective action that generates %
the one-particle irreducible Green's functions \cite{wett93,btw02}.
A regulator is added to the theory to suppress fluctuations with momenta below 
some scale $k$. For large $k$, we start with a suitably parametrised ``bare" action.
The methd works by evolving from this bare action to the limit $k \to 0$, where 
all quantum fluctuations are included and the action becomes physical. Away from
this limit, that is for $k > 0$, the running action is not physical 
because of the partial suppression of fluctuations.
Even though it is fully nonperturbative, the driving 
term in the FRG equation for the action has the form of a one-loop integral. 
Instead of diagrammatic expansions, practical approximation schemes are obtained
by truncating the effective action to a finite number of terms. 

This FRG is being applied to systems of nonrelativistic particles, in order to 
study, in particular, dense matter \cite{bkmw05,dgpw07a,dfgpw10}. In that 
context, it provides an alternative to traditional many-body methods. As 
part of this programme, studies of few-body systems in the same framework 
are needed to fix the input parameters. These studies are also proving 
interesting in their own right \cite{fmsw09,mfsw09,sm10,bkw11}.

A key ingredient of our approach is a trimer field. Such fields have been 
introduced before, in Refs.~\cite{fmsw09,mfsw09,sm10}. However in the
previous application to the four-boson problem \cite{sm10}, this 
field was used to explore the dependence of amplitudes on an external energy.
In contrast, our approach emphasises its dynamical role. This allows us to
describe the atom-trimer channel of the four-body system and hence 
to obtain equations with a structure like that of 
the Faddeev-Yakubovsky equations \cite{yak67}.

In this work we study systems of up to four nonrelativistic bosonic 
``atoms". We represent the atoms by the field $\psi(x)$ and we also introduce 
dimer and trimer fields, $\phi(x)$ and $\chi(x)$, in order to include 
energy-dependent propagators for two- and three-body subsystems. 
The evolution equation for the effective action 
$\Gamma_k[\psi,\psi^*,\phi,\phi^*,\chi,\chi^*]$
takes the form \cite{btw02}
\begin{equation}\label{eq:flow:eq}
\partial_k\Gamma=-\frac{\mbox{i}}{2}\,\Trace \left[(\partial_k{\mathbf R})\,
\left(({\boldsymbol \Gamma}^{(2)}-{\mathbf R})^{-1}\right)\right]
+\frac{\delta\Gamma}{\delta\Phi}\cdot\partial_k\Phi,
\end{equation}
where ${\boldsymbol \Gamma}^{(2)}$ denotes the matrix of second derivatives of 
the action with respect to the fields and ${\mathbf R}$ the regulator that is added 
to suppress low-momentum modes. The trace $\Trace$ and the scalar product in the 
final term include integrals over energy and three-momenta as well as sums over the
different types of field. The final term in the equation appears when we include 
fields that depend explicitly on the scale $k$, as in Ref.~\cite{gw02,fmsw09,sm10}.

For our regulators, ${\mathbf R}$, we use the form suggested by Litim \cite{lit01}, which is
optimised for local interactions. This suppresses the contributions of modes 
with momenta $q < k$ by replacing the kinetic energy 
in the inverse propagator for each field with a constant.
For the atom field it has the form
\begin{equation}\label{eq:regulator}
\ra(q,k)=\frac{k^2-q^2}{2m}\,\theta(k-q).
\end{equation}
The dimer and trimer regulators have similar forms but also contain the 
wave-function renormalisation factors defined below.

The key ingredient in any practical application of the FRG is the choice of 
truncation for the running action. Here we work with only local 
interactions. This reduces the functional differential equation 
for the action to a set of coupled ordinary differential equations for 
renormalisation factors and coupling constants multiplying the terms in
that appear in the action, as defined below.
Large numbers of diagrams contribute to the driving terms, 
as in the versions without trimer fields studied in Refs.~\cite{sm10,bkw11}.

The running action we use is 
\begin{widetext}
\begin{eqnarray}\label{eq:running:action}
\Gamma_k[\psi,\psi^*,\phi,\phi^*,\chi,\chi^*]
&=&\int{\rm d}^4x\,\Biggl[
\psi^*\left({\rm i}\,\partial_0+\frac{\nabla^2}{2m}\right)\psi
+Z_d\,\phi^*\left({\rm i}\,\partial_0+\frac{\nabla^2}{4m}\right)\phi
+Z_t\,\chi^*\left({\rm i}\,\partial_0+\frac{\nabla^2}{6m}\right)\chi\cr
\noalign{\vspace{5pt}}
&&\qquad\qquad-u_d\phi^*\phi-u_t\chi^*\chi
-\frac{g}{2}\bigl(\phi^*\psi\psi+\psi^*\psi^*\phi\bigr)
-h\bigl(\chi^*\phi\psi+\phi^*\psi^*\chi\bigr)-\lambda\,\phi^*\psi^*\phi\psi\cr
\noalign{\vspace{5pt}}
&&\qquad\qquad-\frac{u_{dd}}{2}\bigl(\phi^*\phi\bigr)^2
-\frac{v_d}{4}\bigl(\phi^*\phi^*\phi\psi\psi+\phi^*\psi^*\psi^*\phi\phi\bigr)
-\frac{w}{4}\phi^*\psi^*\psi^*\phi\psi\psi\cr
&&\qquad\qquad\null -u_{tt}\, \chi^*\psi^*\chi\psi
-\frac{u_{dt}}{2}\bigl(\phi^*\phi^*\chi\psi+\chi^*\psi^*\phi\phi\bigr)
-\frac{v_t}{2}\bigl(\phi^*\psi^*\psi^*\chi\psi+\chi^*\psi^*\phi\psi\psi\bigr)
\Biggr].
\end{eqnarray}
\end{widetext}
This contains kinetic terms
for atom, dimer and trimer fields with wave-function renormalisation factors and 
interaction terms with up to four underlying atoms. This action was also used by Schmidt and Moroz \cite{sm10} (see in particular the Appendix to that paper) but
they chose to eliminate the four-atom couplings with trimer fields 
($u_{tt}$, $u_{dt}$ and $v_t$) so that channels with dynamic trimers are not 
needed. The analogous fermionic couplings without trimers were studied in 
Ref.~\cite{bkw11}. 

The inverse propagators for the fields in Eq.~(\ref{eq:running:action}) 
are expanded up to first order in the energy, which impliess first-order time derivatives in the action. In each channel, the zero-energy point for this 
expansion is taken to be the threshold for breakup of an $n$-atom state into 
$n$ free atoms. Spatial derivatives appear at second order as required by 
Galilean invariance, which follows from our choice of regulator \cite{lit01}.

The wave-function renormalisation factors $Z_{d,t}$, self-energies 
$u_{d,t}$ and couplings $h$, $\lambda$ etc.~all run with the regulator scale $k$. 
In vacuum, there is no renormalisation factor for the atom field $\psi$ 
and the coupling $g$ remains constant during the evolution.

Even though atom-atom scattering near the unitary limit can be described by an 
atom-atom contact interaction, the running action in Eq.~(\ref{eq:running:action})
does not contain such a term. This is because it can be eliminated through a 
Hubbard-Stratonovich transformation at some large starting scale $K$.
The atom-atom term is not regenerated by the evolution 
and so atom-atom scattering is mediated only by the coupling 
$g$ to dimers. At zero energy, the scattering is given by $g^2/u_d(k)$ where 
$u_d(k)$ evolves linearly with $k$. We choose its initial value $u_d(K)$ such that, 
in the physical limit, $u_d(0)$ gives the desired scattering length 
$a$ \cite{bkmw05,dgpw07a}.

In contrast, the atom-dimer interaction, $\lambda$, is regenerated even if we 
set it to zero initially. By introducing fields that depend explicitly on the 
scale $k$, as in Ref.~\cite{gw02,fmsw09,sm10}, we can cancel the evolution of this
and some other couplings. If we set their initial values to zero at the starting 
scale $K$, then these couplings are effectively eliminated from the problem.
Here we take the trimer to run as
\begin{equation}\label{eq:dk:chi}
\partial_k \chi = \zeta_1 \, \phi \psi
 + \zeta_2 \, \psi^{\dagger} \chi \psi
+ \zeta_3 \, \psi^{\dagger} \phi \phi
+ \zeta_4 \, \psi^{\dagger} \phi \psi \psi,
\end{equation}
where the $\zeta_i(k)$ are
\begin{subequations} \label{eq:zeta:all}
\begin{eqnarray}
\label{eq:zeta:a}
\zeta_1&=&-\, \frac{\partial_k\lambda}{2 \, h},\\
\noalign{\vspace{5pt}}
\label{eq:zeta:b}
\zeta_2&=& -\, \frac{ \partial_kv_t }{ h }
+\frac{ u_{tt} \, \partial_k\lambda }{ 2 \, h^2 }
- \frac{ u_t \, v_t \, \partial_k\lambda }{ 2 \, h^3 }
+ \frac{ u_t \, \partial_k w }{ 8 \, h^2 },\\
\noalign{\vspace{5pt}}
\label{eq:zeta:c}
\zeta_3&=&-\, \frac{ \partial_kv_d }{ 4 \, h }
+\frac{ u_{dt} \, \partial_k\lambda }{ 2 \, h^2 },
\\
\noalign{\vspace{5pt}}
\label{eq:zeta:d}
\zeta_4&=&-\, \frac{ \partial_kw }{ 8 \, h }
+\frac{ v_t \, \partial_k\lambda}{ 2 \, h^2 }.
\end{eqnarray}
\end{subequations}
The first term in Eq.~(\ref{eq:dk:chi}) cancels the running of $\lambda$,
and the others do the same for the four-atom couplings $v_d$, $w$ and $v_t$.

Once we have eliminated these couplings, the physical processes that give rise 
to their evolution are implicitly present in the flows of the remaining 
couplings through contributions to their flows from the final term in
Eq.~(\ref{eq:flow:eq}). For example, if the contact interaction $\lambda$ 
is eliminated, atom-dimer scattering only occurs through coupling to
the trimer. The effects responsible for the evolution of $\lambda$ are 
now codified in a term proportional to $u_t\zeta_1$ in 
the flow of $h(k)$, arising from the first term of Eq.~(\ref{eq:dk:chi}).

\section{\label{section:three_body}Three-body sector}

The three-body sector, described by the couplings $h(k)$, $u_t(k)$ and $Z_t(k)$, 
has been studied using this action by Floerchinger \textit{et al.}~\cite{fmsw09}. 
We summarise its main features here to provide some ``landmarks" for our 
four-body results.
In the unitary limit, the flow equations for the three-atom couplings have the forms
\begin{subequations}\label{eq:three:body:flows:all}
\begin{eqnarray}
\label{eq:three:body:flows:h}
\partial_k\bigl(h^2\bigr)
&=&-\,\frac{312}{125 \, k} \; h^2(k)
- \frac{256 }{125 \, k^3} \; u_t(k),\\
\noalign{\vspace{5pt}}
\label{eq:three:body:flows:uchi}
\partial_k u_t
&=&\frac{56 \, k}{125 } \, h^2(k),\\
\noalign{\vspace{5pt}}
\label{eq:three:body:flows:zchi}
\partial_k Z_t
&=&-\,\frac{448}{625 \, k} \; h^2(k),
\end{eqnarray}
\end{subequations}
where, to simplify the expressions, $u_t$ and $Z_t$ have been redefined to absorb 
constant factors of $g^2 \, m$ and $1/g^2$ respectively.

These equations describe the flows for regulator scales $k \gg 1/a$, where $a$ is 
the two-body scattering length. In this limit, the equations are scale invariant
and so we expect their solutions to scale as powers of $k$. 
Indeed this system of differential equations is satisfied if 
$h^2(k)$ and $Z_t(k)$ behave as $k^{ d }$ and $u_t(k)$ as $k^{ 2 + d }$ 
where $d$ has two possible values, 
\begin{equation}\label{eq:three:body:exponent:in:h}
d_{\pm} = -281/125 \pm \mathfrak{i} \, \sqrt{535}/25.
\end{equation}
Since these are a complex-conjugate pair, we can form real solutions 
and define rescaled quantities that oscillate periodically in $t = \ln(k/K)$:
\begin{subequations}\label{eq:three:body:scalings:all}
\begin{eqnarray}
\label{eq:three:body:scalings:H}
\hat{H}(k)&=&k^{281/125} \; h^2(k),\\
\noalign{\vspace{5pt}}
\label{eq:three:body:scalings:uchi}
\hat{u}_t(k)&=&k^{31/125} \; u_t(k),\\
\noalign{\vspace{5pt}}
\label{eq:three:body:scalings:zchi}
\hat{Z}_t(k)&=&k^{281/125} \; Z_t(k).
\end{eqnarray}
\end{subequations}

This periodic behaviour is a consequence of the Emimov effect \cite{efim70,efim71}
which breaks the scale invariance of theory to a discrete symmetry. It follows from 
the complex scaling exponents in Eq.~(\ref{eq:three:body:exponent:in:h}).
For the truncated action and regulator used here, the scaling factor in momentum is
$\sim 29.8$ \cite{fmsw09,sm10}, which yields longer cycles than the true value of 
$\sim 22.7$.

In this framework, atom-dimer scattering at zero energy is given by the combination 
$h(k)^2/u_t(k)$, which evolves in the same way as $\lambda(k)$ in the theory 
without the trimer \cite{fmsw09,sm10}. It displays a sequence of poles that are 
equally spaced in $t$, reflecting the discrete scaling symmetry of the Efimov 
effect. Each of these poles corresponds to the passage of a three-body bound 
state through the three-atom threshold as $k$ is lowered. In the physical limit 
they build up the infinite tower of Efimov states.

Although the flow equations in the three-body sector require three initial 
conditions, only one of these defines a physical parameter. This fixes the 
initial phase of the periodic functions or, equivalently, the scale at which 
the first Efimov pole appears. Physical quantities are independent of the 
magnitudes of the couplings since they depend only on the ratios $h(k)^2/u_t(k)$
and $h(k)^2/Z_t(k)$.

\section{\label{section:four_body}Four-body sector}

In the four-atom sector, we use the scale dependence of the trimer to 
eliminate the couplings $v_d$, $w$ and $v_t$ that include the dimer-atom-atom 
channel. This leaves only the ones involving the dimer-dimer and 
atom-trimer channels, $u_{dd}$, $u_{dt}$ and $u_{tt}$. The first of these, 
$u_{dd}$, describes dimer-dimer scattering at zero energy (the four-atom
threshold). Similarly $u_{tt}$ describes atom-trimer scattering and $u_{dt}$
the coupling between the two channels. This choice reflects the structure of the 
Faddeev-Yakubovsky equations used in most direct calculations of four-body 
systems \cite{yak67}. In contrast, Schmidt and Moroz \cite{sm10} also 
introduced a trimer field to treat energy dependence but kept 
only the couplings $u_{dd}$, $v_d$ and $w$.

The evolution of the four-atom couplings, $u_{dd}$, $u_{dt}$ and $u_{tt}$,
is governed by a system of three coupled nonlinear differential equations.
We define regulated energies for atoms, dimers and trimers, 
\begin{subequations}\label{eq:regulated:energies:all}
\begin{eqnarray}
\ea(q,k) &=& \frac{q^2}{2 \, m} + \ra(q,k),
\label{eq:regulated:energy:atom}\\
\noalign{\vspace{5pt}}
\ed(q,k) &=& \frac{q^2}{4 \, m} + \frac{\rd(q,k)}{Z_d(k)} + \frac{u_d(k)}{Z_d(k)},
\label{eq:regulated:energy:dimer}\\
\noalign{\vspace{5pt}}
\et(q,k) &=& \frac{q^2}{6 \, m} + \frac{\rt(q,k)}{Z_t(k)} + \frac{u_t(k)}{Z_t(k)},
\label{eq:regulated:energy:trimer}
\end{eqnarray}
\end{subequations}
where the single-atom self-energy contains
\begin{equation}
u_d(k,a) = \frac{M \, g^2}{\pi^2} \; \left( \frac{k}{6} - \frac{\pi}{4 \, a} \right).
\end{equation}
From these we construct the quantities 
\begin{eqnarray}
\tgen{\alpha}{\beta}{\gamma}{\delta}{X}
=\frac{\partial_k R_X \; \left( \zphi \right)^{-\beta-\gamma} 
\left( \zchi \right)^{-\delta} }
{ \left( \ea \right)^\alpha \left( \ed \right)^\beta \left( \ea + \ed \right)^\gamma 
\left( \ea + \et \right)^\delta },
\end{eqnarray}
for $X = a,d,t$. In terms of these, the system of equations can be written
\begin{widetext}
\begin{subequations}\label{eq:fbp:all}
\begin{eqnarray}
\partial_k u_{dd}&=&\int \! \frac{d^3q}{\left( 2 \, \pi \right)^3}
\bigg[\frac{3 \, g^4}{8} \tatom{4}{0}{0}{0}
+ \frac{g^2 h^2}{2} \left( 2 \, \tatom{3}{0}{0}{1} + \tatom{2}{0}{0}{2} \zchi 
+ \ttrimer{2}{0}{0}{2} \right)
+ \frac{\left( u_{dd} \right)^2}{2} \tdimer{0}{2}{0}{0} \left( \zphi \right)^{-1}
\nonumber \\
&&\qquad\qquad\quad-\, 2 \, g \, h \, u_{dt} \left( \tatom{2}{0}{0}{1} 
+ \tatom{1}{0}{0}{2} \zchi + \ttrimer{1}{0}{0}{2} \right)
+ 2 \left( u_{dt} \right)^2 \left( \tatom{0}{0}{0}{2} \zchi 
+ \ttrimer{0}{0}{0}{2} \right)\bigg],
\end{eqnarray}
\begin{eqnarray}
\partial_k u_{dt}&=&\int \! \frac{d^3q}{\left( 2 \, \pi \right)^3}
\bigg[- \frac{g^3 u_t \, u_{dd}}{4 \, h} \left( \tatom{2}{1}{1}{0} + \tatom{1}{1}{2}{0} \zphi + \tdimer{1}{2}{1}{0} + \tdimer{1}{1}{2}{0} \right)
- \frac{g \, h \, u_{tt}}{2} \left( \tatom{2}{0}{0}{1} + \tatom{1}{0}{0}{2} \zchi 
+ \ttrimer{1}{0}{0}{2} \right)\nonumber \\
&&\qquad\qquad\quad
-\, \frac{g \, h \, u_{dd}}{2} \left( \tatom{0}{1}{2}{0} \zphi + \tdimer{0}{2}{1}{0} 
+ \tdimer{0}{1}{2}{0} \right)
+ \frac{g^4 u_t \, u_{dt}}{8 \, h^2} \left( 2 \, \tatom{3}{0}{1}{0} 
+ \tatom{2}{0}{2}{0} \zphi + \tdimer{2}{0}{2}{0} \right)\nonumber \\
&&\qquad\qquad\quad
-\, \frac{g^2 u_t \, u_{dt}}{2} \left( \tatom{2}{0}{1}{1} + \tatom{1}{0}{2}{1} \zphi 
+ \tatom{1}{0}{1}{2} \zchi + \tdimer{1}{0}{2}{1} + \ttrimer{1}{0}{1}{2} \right)
+ \frac{u_{dd} \, u_{dt}}{2} \tdimer{0}{2}{0}{0} \left( \zphi \right)^{-1}
\nonumber \\
&&\qquad\qquad\quad
-\, h^2 u_{dt} \left( \tatom{0}{0}{2}{1} \zphi + \tatom{0}{0}{1}{2} \zchi + \tdimer{0}{0}{2}{1} + \ttrimer{0}{0}{1}{2} \right)
+ u_{dt} \, u_{tt} \left( \tatom{0}{0}{0}{2} \zchi + \ttrimer{0}{0}{0}{2} \right)
\nonumber \\
&&\qquad\qquad\quad
+\, \frac{g^3 h \, u_t}{4} \left( 2 \, \tatom{3}{0}{1}{1} + \tatom{2}{0}{2}{1} \zphi 
+ \tatom{2}{0}{1}{2} \zchi + \tdimer{2}{0}{2}{1} + \ttrimer{2}{0}{1}{2} \right)
\nonumber \\
&&\qquad\qquad\quad
+\, \frac{g \, h^3}{2} \left( \tatom{2}{0}{1}{1} + \tatom{1}{0}{2}{1} \zphi + \tatom{1}{0}{1}{2} \zchi + \tdimer{1}{0}{2}{1} + \ttrimer{1}{0}{1}{2} \right)\bigg],
\end{eqnarray}
\begin{eqnarray}
\partial_k u_{tt}&=&\int \! \frac{d^3q}{\left( 2 \, \pi \right)^3}
\bigg[ g^2 h^2 \left( \tatom{2}{1}{1}{0} + \tatom{1}{1}{2}{0} \zphi 
+ \tdimer{1}{2}{1}{0} + \tdimer{1}{1}{2}{0} \right)
+ \frac{g^4 u_t \, u_{tt}}{4 \, h^2} \left( 2 \, \tatom{3}{0}{1}{0} 
+ \tatom{2}{0}{2}{0} \zphi + \tdimer{2}{0}{2}{0} \right)
\nonumber \\
&&\qquad\qquad\quad
+\, g^4 \, u_t \left( 2 \, \tatom{3}{1}{1}{0} + \tatom{2}{1}{2}{0} \zphi 
+ \tdimer{2}{2}{1}{0} + \tdimer{2}{1}{2}{0} \right)
- 2 \, g \, h \, u_{dt} \left( \tatom{0}{1}{2}{0} \zphi 
+ \tdimer{0}{2}{1}{0} + \tdimer{0}{1}{2}{0} \right)
\nonumber \\
&&\qquad\qquad\quad
+\, \frac{g^4 \left( u_t \right)^2}{4} \left( 2 \, \tatom{3}{0}{2}{1} 
+ 2 \, \tatom{2}{0}{3}{1} \zphi + \tatom{2}{0}{2}{2} \zchi 
+ 2 \, \tdimer{2}{0}{3}{1} + \ttrimer{2}{0}{2}{2} \right)
+ \left( u_{dt} \right)^2 \tdimer{0}{2}{0}{0} \left( \zphi \right)^{-1}
\nonumber \\
&&\qquad\qquad\quad
-\, 2 \, h^2 u_{tt} \left( \tatom{0}{0}{2}{1} \zphi + \tatom{0}{0}{1}{2} \zchi 
+  \tdimer{0}{0}{2}{1} + \ttrimer{0}{0}{1}{2} \right)
+ \left( u_{tt} \right)^2 \left( \tatom{0}{0}{0}{2} \zchi 
+ \ttrimer{0}{0}{0}{2} \right)
\nonumber \\
&&\qquad\qquad\quad
+\, g^2 h^2 u_t \left( \tatom{2}{0}{2}{1} + 2 \, \tatom{1}{0}{3}{1} \zphi 
+ \tatom{1}{0}{2}{2} \zchi + 2 \, \tdimer{1}{0}{3}{1} 
+ \ttrimer{1}{0}{2}{2} \right)
- g^2 u_t \, u_{tt} \left( \tatom{2}{0}{1}{1} + \tatom{1}{0}{2}{1} \zphi \right)
\nonumber \\
&&\qquad\qquad\quad
-\, g^2 u_t \, u_{tt} \left( \tatom{1}{0}{1}{2} \zchi + \tdimer{1}{0}{2}{1} 
+ \ttrimer{1}{0}{1}{2} \right)
+ \frac{g^6 \left( u_t \right)^2}{4 \, h^2} \left( 3 \, \tatom{4}{1}{1}{0} 
+ \tatom{3}{1}{2}{0} \zphi + \tdimer{3}{2}{1}{0} + \tdimer{3}{1}{2}{0} \right)
\nonumber \\
&&\qquad\qquad\quad
-\, \frac{g^3 u_t \, u_{dt}}{h} \left( \tatom{2}{1}{1}{0} + \tatom{1}{1}{2}{0} \zphi 
+ \tdimer{1}{2}{1}{0} + \tdimer{1}{1}{2}{0} \right)
\nonumber \\
&&\qquad\qquad\quad
+\, h^4 \left( 2 \, \tatom{0}{0}{3}{1} \zphi + \tatom{0}{0}{2}{2} \zchi 
+ 2 \, \tdimer{0}{0}{3}{1} + \ttrimer{0}{0}{2}{2} \right)\bigg].
\end{eqnarray}
\end{subequations}
\end{widetext}

The appearance of $h^2(k)$, $u_t(k)$ and $Z_t(k)$ in the four-body flow equations, 
Eqs.~(\ref{eq:fbp:all}), means that they inherit the Efimov periodicity of the 
three-body sector. This also leads to two types of singularity in the equations.
One arises from terms with denominators containing either one or two powers of 
the regulated energy of an atom plus a trimer, $\ea(k) + \et(k)$. This passes 
through zero energy once in every Efimov cycle, at the point where a regulated 
atom-trimer threshold drops below the four-atom threshold as we lower 
$k$. At each crossing we expect additional contributions to the imaginary parts of 
the four-body couplings, as a channel with a new Efimov state becomes open.

The other type of divergent term has a factor of $1/\left(h(k)\right)^2$.
These lead to unphysical singularities in the four-body couplings, which mark 
the start of a short region within each Efimov cycle where $h^2(k)$ and $Z_t(k)$ 
have opposite signs. In these regions, the trimer field has a ghost-like character, 
with a propagator $h^2(k)/(Z_t\,p_0-u_t(k))$ that has a negative residue at its pole.
This is a warning that not all features of the effective action are physical 
for non-zero values of $k$. Fortunately these regions are well separated from the 
threshold regions where the phenomena of interest occur.

In the scaling regime the four-atom couplings display Efimov periodicity. This 
can be seen most clearly if they are multiplied by appropriate powers of $k$, 
analogously to the rescaling of the three-body sector in Eqs.~(\ref{eq:three:body:scalings:all}).
Here we define the couplingss, 
\begin{subequations}\label{eq:fbp:scalings:all}
\begin{eqnarray}
\hat{u}_{dd}(k)&=&k^3 \;\, u_{dd}(k),
\label{eq:fbp:scalings:udd}\\
\hat{u}_{dt}(k)&=&k^{781/250} \;\, u_{dt}(k),
\label{eq:fbp:scalings:udt}\\
\hat{u}_{tt}(k)&=&k^{406/125} \;\, u_{tt}(k),
\label{eq:fbp:scalings:utt}
\end{eqnarray}
\end{subequations}
where the powers of $k$ are determined from dimensional analysis of the running 
action and the scalings in the three-body sector, 
Eqs.~(\ref{eq:three:body:flows:all}).

The flow equations for these rescaled couplings can be written
\begin{widetext}
\begin{subequations} \label{eq:flow:all}
\begin{eqnarray}
\label{eq:flow:uu2}
\partial_t \hat{u}_{dd}&=&  \frac{1}{\pi^2}+ 3 \, \hat{u}_{dd}
+ \frac{8\pi^2 \, \hat{u}_{dd}^2}{15}
+ \frac{\hat{H} \; \partial_t \hat{Z}_t}{45\pi^2 \, \hat E_{at}^2 \; \hat{Z}_t^2}
+ \frac{2 \, \hat{H} \; \hat{u}_t}{3\pi^2 \, \hat E_{at}^2 \hat{Z}_t^2}
+ \frac{1573 \, \hat{H}}{1875\pi^2 \, \hat E_{at}^2 \; \hat{Z}_t}
+ \frac{2 \, \hat{H}}{3\pi^2 \, \hat E_{at} \hat{Z}_t}
- \frac{2 \, \hat{U}_{dt} \; \partial_t \hat{Z}_t}{45\pi^2 \, \hat E_{at}^2 
\; \hat{Z}_t^2}\nonumber \\
\noalign{\vspace{5pt}}
&&-\, \frac{2 \, \hat{u}_t \; \hat{U}_{dt}}{3\pi^2 \, \hat E_{at}^2 \; \hat{Z}_t^2}
- \frac{6938 \, \hat{U}_{dt}}{5625\pi^2 \, \hat E_{at}^2 \; \hat{Z}_t}
- \frac{2 \, \hat{U}_{dt}}{3\pi^2 \, \hat E_{at} \; \hat{Z}_t}
+ \frac{\hat{U}_{dt}^2 \; \partial_t \hat{Z}_t}{45\pi^2 \, \hat{H} \; 
\hat E_{at}^2 \; \hat{Z}_t^2}
+ \frac{2219 \, \hat{U}_{dt}^2}{5625\pi^2 \, \hat{H} \; \hat E_{at}^2 \; \hat{Z}_t},
\end{eqnarray}
\begin{eqnarray}
\label{eq:flow:h:times:udt}
\partial_t  \hat{U}_{dt} &=&-\, \frac{496\pi^2 \, \hat{H} \; \hat{u}_{dd}}{375}
- \frac{1096\pi^2 \, \hat{u}_{dd} \; \hat{u}_t}{375}
+ \frac{2 \, \hat{H}^2 \; \partial_t \hat{Z}_t}{75 \, \hat E_{at}^2 \; \hat{Z}_t^2}
+ \frac{2 \, \hat{H} \; \hat{u}_t \; \partial_t \hat{Z}_t}{75 \, \hat E_{at}^2 \; 
\hat{Z}_t^2}
+ \frac{18 \, \hat{H}^2 \; \hat{u}_t}{25 \, \hat E_{at}^2 \hat{Z}_t^2}
+ \frac{28 \, \hat{H} \; \hat{u}_t^2}{25 \, \hat E_{at}^2 \; \hat{Z}_t^2}
+ \frac{8938 \, \hat{H}^2}{9375 \, \hat E_{at}^2 \; \hat{Z}_t}
\nonumber \\
\noalign{\vspace{5pt}}
&&+\, \frac{11438 \, \hat{H} \; \hat{u}_t}{9375 \, \hat E_{at}^2 \; \hat{Z}_t}
+ \frac{66 \, \hat{H}^2}{125 \, \hat E_{at} \; \hat{Z}_t}
+ \frac{116 \, \hat{H} \; \hat{u}_t}{125 \, \hat E_{at} \; \hat{Z}_t}
- \frac{\hat{H} \; \hat{u}_{tt} \; \partial_t \hat{Z}_t}{90 \pi^2 \, \hat E_{at}^2 \; \hat{Z}_t^2}
- \frac{\hat{H} \; \hat{u}_t \; \hat{u}_{tt}}{6 \pi^2 \, \hat E_{at}^2 \; \hat{Z}_t^2}
- \frac{3469 \, \hat{H} \; \hat{u}_{tt}}{11250 \pi^2 \, \hat E_{at}^2 \; \hat{Z}_t}
\nonumber \\
\noalign{\vspace{5pt}}
&&-\, \frac{\hat{H} \; \hat{u}_{tt}}{6 \pi^2 \, \hat E_{at} \; \hat{Z}_t}
+ 3 \, \hat{U}_{dt}+ \frac{8\pi^2 \, \hat{u}_{dd} \; \hat{U}_{dt}}{15}
- \frac{2 \, \hat{H} \; \hat{U}_{dt} \; \partial_t \hat{Z}_t}{75 \, \hat E_{at}^2 \; \hat{Z}_t^2}
- \frac{2 \, \hat{u}_t \; \hat{U}_{dt} \; \partial_t \hat{Z}_t}{75 \, \hat E_{at}^2 \; \hat{Z}_t^2}
- \frac{8 \, \hat{H} \; \hat{u}_t \; \hat{U}_{dt}}{25 \, \hat E_{at}^2 \; \hat{Z}_t^2}
- \frac{18 \, \hat{u}_t^2 \; \hat{U}_{dt}}{25 \, \hat E_{at}^2 \; \hat{Z}_t^2}
\nonumber \\
\noalign{\vspace{5pt}}
&&-\, \frac{2146 \, \hat{H} \; \hat{U}_{dt}}{3125 \, \hat E_{at}^2 \; \hat{Z}_t}
- \frac{8938 \, \hat{u}_t \; \hat{U}_{dt}}{9375 \, \hat E_{at}^2 \; \hat{Z}_t}
- \frac{16 \, \hat{H} \; \hat{U}_{dt}}{125 \, \hat E_{at} \; \hat{Z}_t}
- \frac{66 \, \hat{u}_t \; \hat{U}_{dt}}{125 \, \hat E_{at} \; \hat{Z}_t}
+ \frac{\hat{u}_{tt} \; \hat{U}_{dt} \; \partial_t \hat{Z}_t}{90 \pi^2 \, \hat E_{at}^2 \; \hat{Z}_t^2}
+ \frac{2219 \, \hat{u}_{tt} \; \hat{U}_{dt}}{11250 \pi^2 \, \hat E_{at}^2 \; \hat{Z}_t},
\end{eqnarray}
\begin{eqnarray}
\label{eq:flow:utt}
\partial_t  \hat{u}_{tt} 
&=& \frac{4384 \pi^2 \, \hat{H}}{375}
+ \frac{13568 \pi^2 \, \hat{u}_t}{375}
+ \frac{9184 \pi^2 \,\hat{u}_t^2}{375 \, \hat{H}}
+ \frac{8 \pi^2 \, \hat{H}^2 \; \partial_t \hat{Z}_t}{125 \, \hat E_{at}^2 \; \hat{Z}_t^2}
+ \frac{16 \pi^2 \, \hat{H} \; \hat{u}_t \; \partial_t \hat{Z}_t}{125 \, \hat E_{at}^2 \; \hat{Z}_t^2}
+ \frac{192 \pi^2 \, \hat{H}^2 \; \hat{u}_t}{125 \, \hat E_{at}^2 \; \hat{Z}_t^2}
+ \frac{8 \pi^2 \, \hat{u}_t^2 \; \partial_t \hat{Z}_t}{125 \, \hat E_{at}^2 \hat{Z}_t^2}
\nonumber \\
\noalign{\vspace{5pt}}
&&+\, \frac{624 \pi^2 \, \hat{H} \; \hat{u}_t^2}{125 \, \hat E_{at}^2 \; \hat{Z}_t^2}
+ \frac{432 \pi^2 \, \hat{u}_t^3}{125 \, \hat E_{at}^2 \; \hat{Z}_t^2}
+ \frac{33752 \pi^2 \, \hat{H}^2}{15625 \, \hat E_{at}^2 \, \hat{Z}_t}
+ \frac{87504 \pi^2 \, \hat{H} \; \hat{u}_t}{15625 \, \hat E_{at}^2 \; \hat{Z}_t}
+ \frac{53752 \pi^2 \; \hat{u}_t^2}{15625 \, \hat E_{at}^2 \; \hat{Z}_t}
+ \frac{384 \pi^2 \, \hat{H}^2}{625 \, \hat E_{at} \; \hat{Z}_t}
\nonumber \\
\noalign{\vspace{5pt}}
&&+\, \frac{1968 \pi^2 \, \hat{H} \; \hat{u}_t}{625 \, \hat E_{at} \; \hat{Z}_t}
+ \frac{1584 \pi^2 \, \hat{u}_t^2}{625 \, \hat E_{at} \; \hat{Z}_t}
+ \frac{406 \, \hat{u}_{tt}}{125}
+ \frac{256 \, \hat{u}_t \; \hat{u}_{tt}}{125 \, \hat{H}}
- \frac{4 \, \hat{H} \; \hat{u}_{tt} \; \partial_t \hat{Z}_t}{75 \, \hat E_{at}^2 \; \hat{Z}_t^2}
- \frac{4 \, \hat{u}_t \; \hat{u}_{tt} \; \partial_t \hat{Z}_t}{75 \, \hat E_{at}^2 \; \hat{Z}_t^2}
- \frac{16 \, \hat{H} \; \hat{u}_t \; \hat{u}_{tt}}{25 \, \hat E_{at}^2 \; \hat{Z}_t^2}
\nonumber \\
\noalign{\vspace{5pt}}
&&-\, \frac{36 \, \hat{u}_t^2 \; \hat{u}_{tt}}{25 \, \hat E_{at}^2 \; \hat{Z}_t^2}
- \frac{4292 \, \hat{H} \; \hat{u}_{tt}}{3125 \, \hat E_{at}^2 \; \hat{Z}_t}
- \frac{17876 \, \hat{u}_t \; \hat{u}_{tt}}{9375 \, \hat E_{at}^2 \; \hat{Z}_t}
- \frac{32 \, \hat{H} \; \hat{u}_{tt}}{125 \, \hat E_{at} \; \hat{Z}_t}
- \frac{132 \, \hat{u}_t \; \hat{u}_{tt}}{125 \, \hat E_{at} \; \hat{Z}_t}
+ \frac{\hat{u}_{tt}^2 \; \partial_t \hat{Z}_t}{90 \pi^2 \, \hat E_{at}^2 \; \hat{Z}_t^2}
+ \frac{2219 \, \hat{u}_{tt}^2}{11250 \pi^2 \hat E_{at}^2 \; \hat{Z}_t}
\nonumber \\
\noalign{\vspace{5pt}}
&&-\, \frac{1984 \pi^2 \, \hat{U}_{dt}}{375}
- \frac{4384 \pi^2 \, \hat{u}_t \; \hat{U}_{dt}}{375 \, \hat{H}}
+ \frac{16 \pi^2 \; \hat{U}_{dt}^2}{15 \, \hat{H}},
\end{eqnarray}
\end{subequations}
\end{widetext}
where we have defined the rescaled atom-trimer energy 
$\hat E_{at} = 2/3 + \hat{u}_\chi/\hat{Z}_\chi$ and the modified coupling
$\hat{U}_{dt} = \hat{H}^{1/2} \; \hat{u}_{dt}$. As in 
Eqs.~(\ref{eq:three:body:flows:all}) we have absorbed powers of the 
constants $g^2$ and $m$ into the couplings to try to simplify the expressions.

\begin{figure} [ht]
\includegraphics[width=\columnwidth]{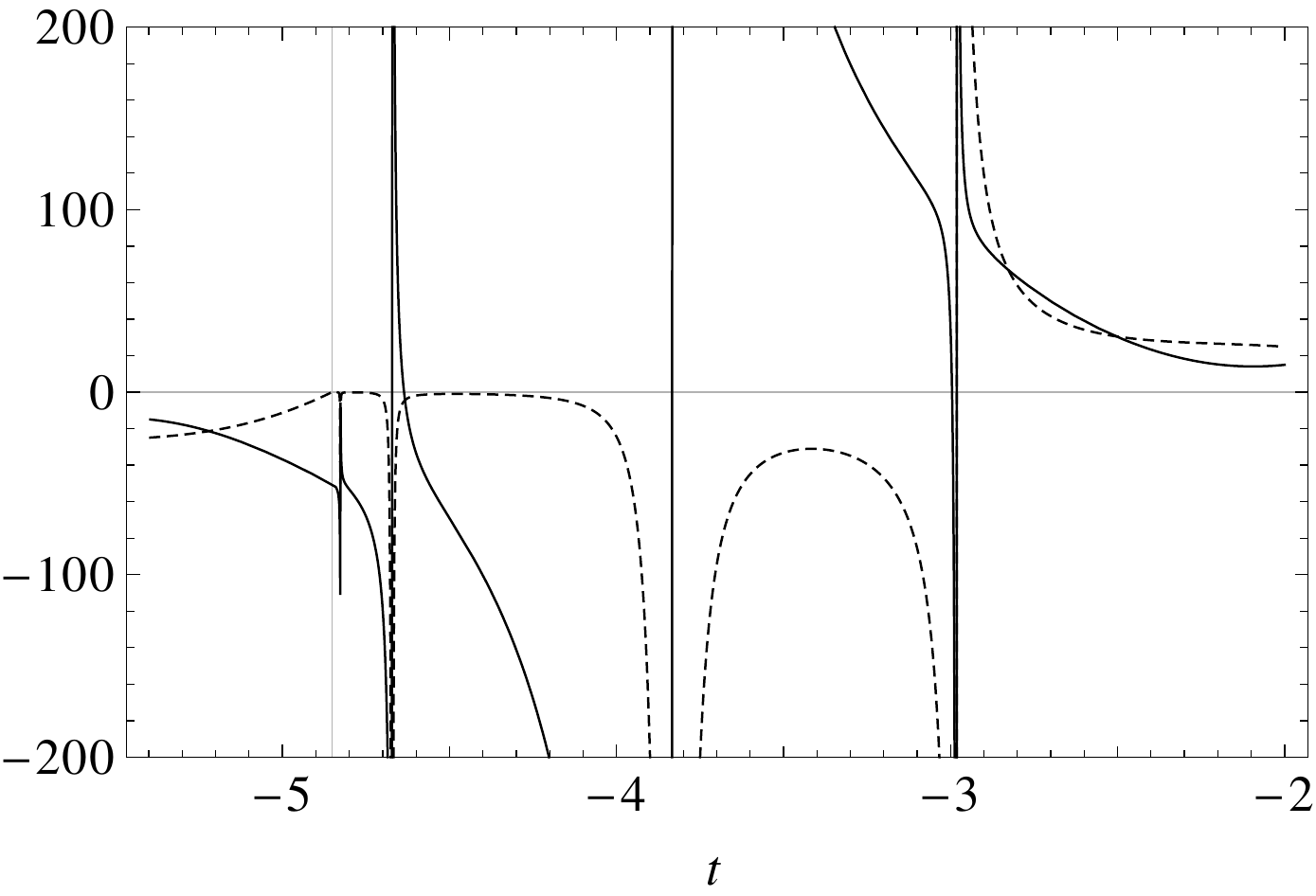}
\caption{
\label{fig:one_cycle}
One Efimov cycle of the flow of the rescaled coupling $\hat{u}_{tt}(k)$ in the
unitary limit, plotted against $t=\ln(k/K)$.
The real part is shown by the solid curve and the imaginary part by the dashed one.
The atom-trimer threshold corresponding to the vanishing of 
$\ea(k) + \et(k)$ is marked by the grey vertical line at 
$t=t_3\simeq-4.85$.}
\end{figure}

We have numerically integrated the coupled equations for $u_{dd}(k)$, $u_{dt}(k)$ 
and $u_{tt}(k)$ through several Efimov cycles, and we have checked that any 
transients caused by our choice of initial conditions die out within the first 
cycle. All three couplings show similar structures but they are most clearly visible 
in $u_{tt}(k)$ and so we present only results for its flow. One cycle of the 
rescaled coupling $\hat{u}_{tt}(k)$ in the unitary limit is shown 
in Fig.~\ref{fig:one_cycle}. At the value of $t=\ln(k/K)$ where the atom-trimer 
threshold passes through zero energy, $t=t_3\simeq -4.85$, we see the expected 
discontinuity in the slope of the imaginary part signalling the opening of a new 
channel. The unphysical singularity arising from the zero of $h^2(k)$ is the 
structure that can be seen at $t\simeq-3.0$. 

\begin{figure} [h]
\includegraphics[width=\columnwidth]{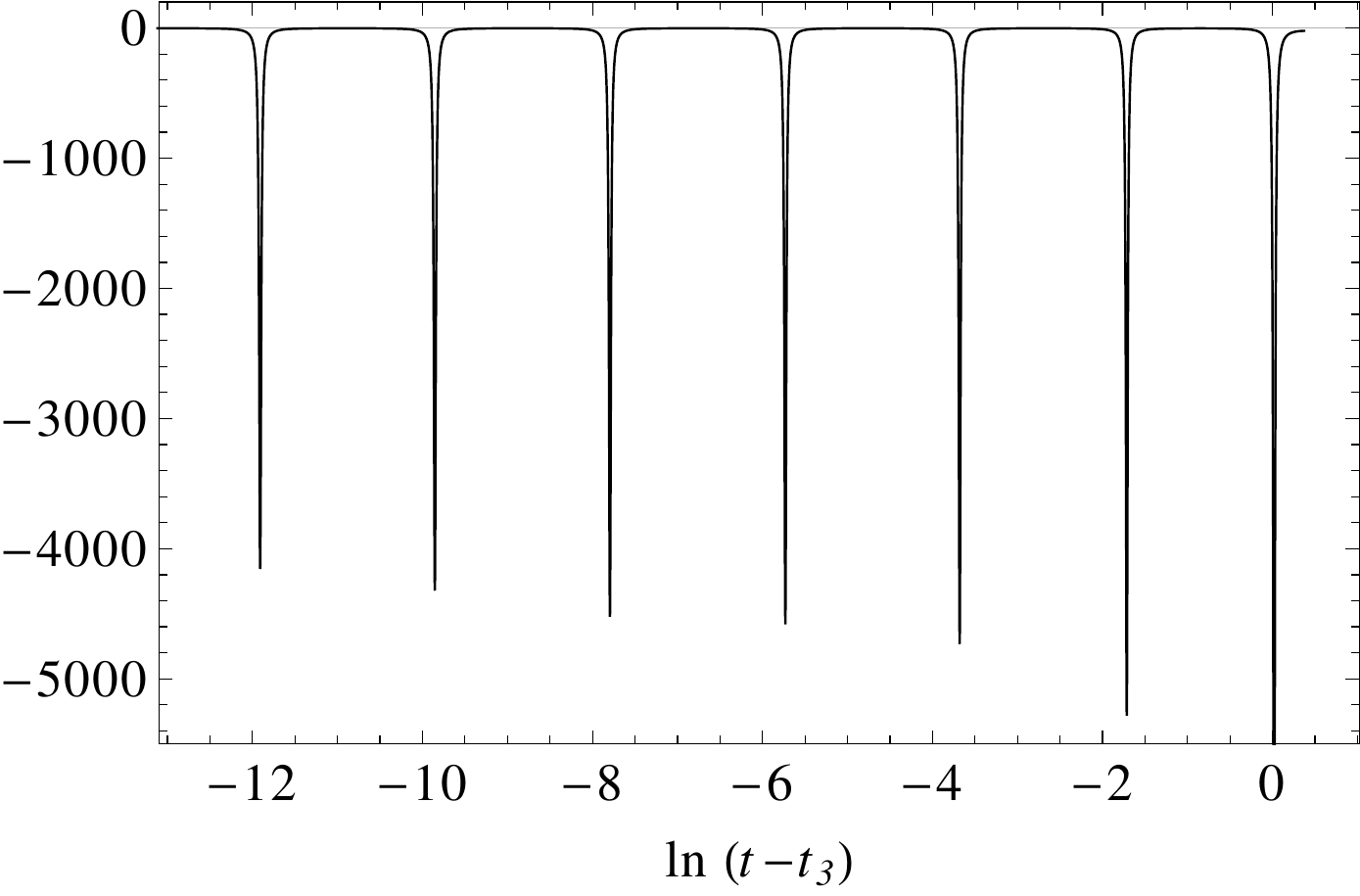}
\caption{\label{fig:below_threshold}
The imaginary part of $\hat{u}_{tt}(k) \, /\left( t - t_3 \right)$ 
just before the threshold $t_3\simeq-4.85$  shown in Fig.~\ref{fig:one_cycle}, 
plotted against $x = \ln \left( t - t_3 \right)$.
Apart from the rightmost one, corresponding to the deepest four-body state, 
the poles are approximately equally spaced.}
\end{figure}

Several simple poles can also be seen in Fig.~\ref{fig:one_cycle}, at 
$t\simeq-3.83$, $-4.67$, and just below the threshold. When we look more closely at 
the region close to an atom-trimer threshold, as in Fig.~\ref{fig:below_threshold}, 
we find an infinite sequence of these poles. These become equally spaced in 
the variable $x=\ln(t-t_3)$. These poles do not correspond to singularities in 
the equations but are generated by the evolution of the couplings.
Like the singularities that appear in the three-body sector, we interpret them as 
bound states or, rather, narrow resonances since they have finite imaginary parts 
as a result of coupling to open channels with more deeply bound trimers.
However, as we discuss below, not all of these poles may appear as 
physical states.

The introduction of the trimer field to describe energy dependence in the 
three-body sector is essential for generating these poles as they do not 
appear in the FRG equations for the couplings without trimer fields \cite{bkw13}.\footnote{The four-body states seen in Ref.~\cite{sm10} have been found to be 
numerical artefacts \cite{bkw13,smpc}.}
The scales at which these poles appear follow a double-exponential, 
``super-Efimov" pattern, similar to that observed in the two-dimensional 
three-body system studied by Nishida \textit{et al.}~\cite{mns13}.

Mathematically this structure arises from the forms of our differential equations 
which are analogous to that of the RG equation of Ref.~\cite{mns13}. The key terms 
that lead to the ``super-Efimov" behaviour are the ones that are singular at the 
atom-trimer threshold. These arise from diagrams that are similar to those in 
Fig.~2 of that paper. However we should stress these states appear for non-zero 
values of $k$, where the action is not physical. Moreover the four-body flow 
equations depend on a scale as a result of the breaking of scale invariance by 
the Efimov effect. These states can therefore move relative to the atom-trimer 
threshold during the evolution to the physical limit. In particular, they may 
pass through the nearby atom-trimer threshold to become virtual states.
If so, only a finite number of bound states may persist in that limit.
Furthermore, a theorem of Amado and Greenwood forbids an infinite number of 
four-body bound states based on a zero-energy trimer state \cite{ag73}.
Nonetheless, the presence of these virtual states might be relevant to the 
rich structure of states being found in for body systems away from the unitary 
limit. For example, Deltuva \cite{delt12} has recently described a tower of 
four-body bound states lying just below the atom-atom-dimer
threshold in systems with finite dimer binding energy.

The local form of the action, Eq.~(\ref{eq:running:action}), does not allow us to 
study the full energy dependence in the four-body channels and so we cannot directly
determine the spectrum in the physical limit. Instead, we can examine where these 
states cross zero energy as we move away from the unitary limit by taking a non-zero 
atom-atom scattering length, $a<0$. Such zero-energy states are the ones
observed in experiments on ultra-cold atoms in traps, as they lead to resonant
enhancements of the loss of atoms at particular values of the scattering length
\cite{fzbhng11,f09}.

\begin{figure} [t]
\includegraphics[width=\columnwidth]{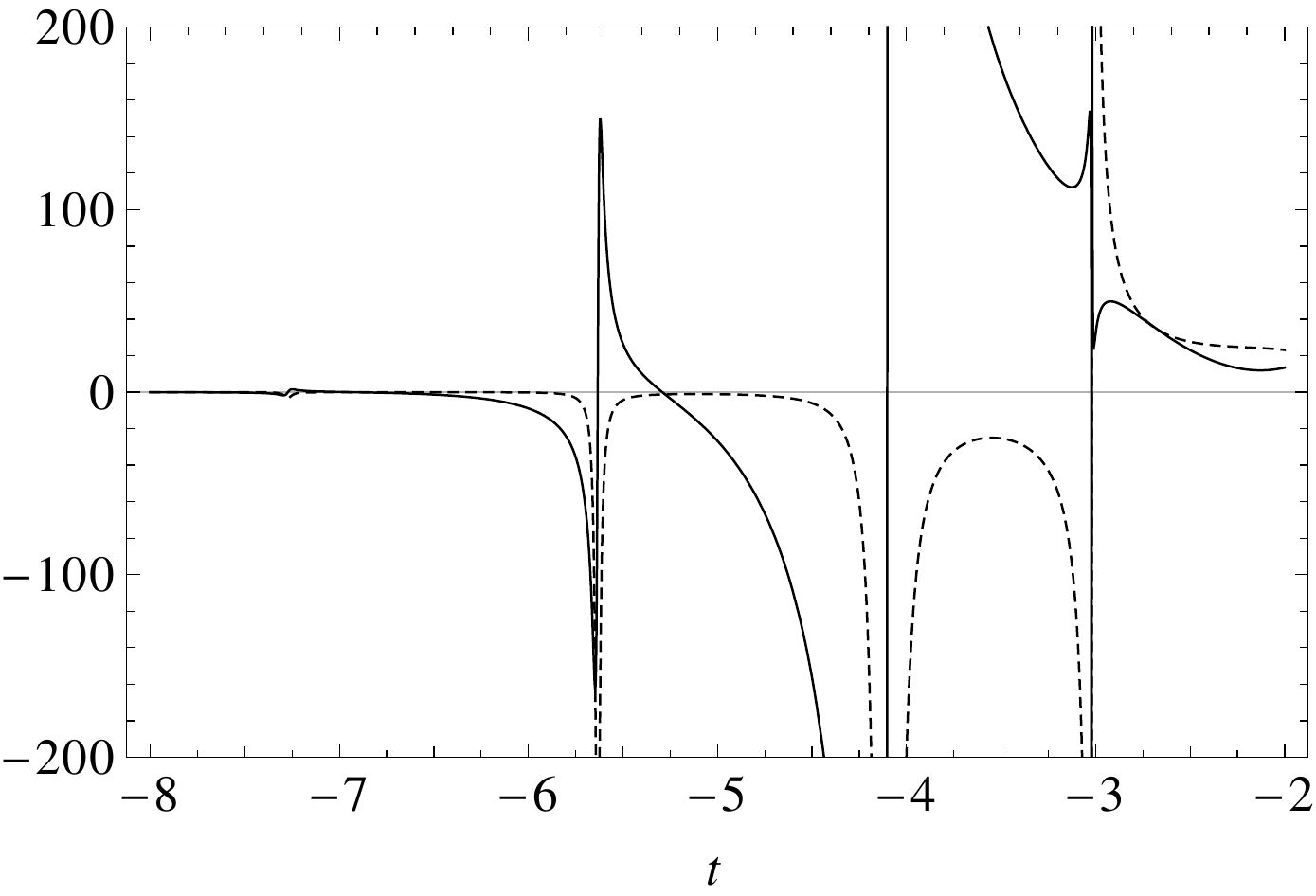}
\caption{\label{fig:last_cycle}
The final cycle of the flow of the rescaled coupling $\hat{u}_{tt}(k)$ 
plotted against $t=\ln(k/K)$.
The solid line corresponds to the real part and the imaginary to the dashed one.
The atom-atom scattering length has been tuned so 
that the last three-body state appears at $k=0$ ($t=-\infty$).}
\end{figure}

With a finite scattering length, the final Efimov cycle no longer has the same 
form as in the unitary limit. An example is shown in Fig.~\ref{fig:last_cycle}.
For $t\gtrsim -2.3$ the flow of the four-atom coupling matches 
Fig.~\ref{fig:one_cycle}, but beyond this point differences become increasingly 
visible. The example shown has the scattering length tuned so that the shallowest 
trimer state has exactly zero binding energy at $k=0$. In this case, we find three 
four-body states appearing in the final Efimov cycle (the poles close to $t=-4.1$, 
$-5.6$ and $-7.1$). There is thus no conflict with the theorem of Amado and Greenwood 
\cite{ag73} that there are only a finite number of these four-body states. We 
denote the corresponding scattering length by $a_3$. When we further decrease $a$, 
we find that the values $a_4^{(n)}$ at which these states cross the four-atom
threshold are related to $a_3$ by
\begin{equation}
a_4^{(0)}\!/a_3\simeq 0.438,\quad
a_4^{(1)}\!/a_3\simeq 0.877,\quad
a_4^{(2)}\!/a_3\simeq 0.9967.
\end{equation}
For the two lowest states, these ratios are within 5\% of the results of exact 
solutions to the four-body equations \cite{vsdg09,delt10}, and hence they are also 
in reasonable agreement with the experimental numbers \cite{f09}. The third state 
lies extremely close to the atom-trimer threshold. If it is real, then it will be 
a challenge to observe both numerically and experimentally. However this state 
may just be an artefact of our truncation since improvements 
to the action which shorten the Efimov cycle might make it unbound.

Returning to the double-exponential behaviour observed during the evolution, 
the scale $k_4^{(n)}$ at which the $n$-th excited four-body state appears 
can be written in the form 
\begin{equation}
k_4^{(n)}=k_3\,\exp\left[\alpha\,{\rm e}^{-\beta n}\right],
\label{eq:k4tok3}
\end{equation}
where $\alpha\simeq 1.53$, $\beta\simeq 2.06$, and $k_3$ denotes the 
scale corresponding to the atom-trimer threshold for the next three-body 
Efimov state.
This describes the energies of all states except the lowest ($n=0$) to a very good approximation.
The ratios between scales for subsequent states can be expressed in the form of a universal scaling function,
\begin{equation}
k_4^{(n+1)}/k_4^{(n)}=\left(k_3/k_4^{(n)}
\right)^{1-\exp(-\beta)}.
\end{equation}
A similar scaling relation between the binding energies has also been found 
by Hadizadeh \textit{et al.}~\cite{had11,had12}, although its functional 
form is quite different and it predicts at most three four-body states in an 
Efimov cycle. More importantly, and in contrast to the results of those 
authors, the scales at which our states appear do not depend on any new four-body 
scale: the parameter $\alpha$ in Eq.(\ref{eq:k4tok3}) 
has a fixed value which is independent of the initial conditions we impose on the 
four-body couplings. The independence of any four-body parameter also applies to 
the physical states discussed above.

\section{\label{section:conclusions}Conclusions}
In summary: we have used the FRG to study systems of four bosons close to the 
unitary limit. In contrast to previous approaches, we introduce a 
dynamical trimer field and use this to match the
channel structure of the Faddeev-Yakubovsky equations. In the physical limit,
where the cut-off scale tends to zero,
we examine the points at which three- and four-body states pass through zero
energy as we vary the atom-atom scattering length. We find three four-body 
states in the last Efimov cycle. The lowest two
of these pass through zero for scattering lengths that are in good
agreement with the results of exact solutions of the Faddeev-Yakubovsky 
equations \cite{vsdg09,delt10} and with experimental observations \cite{f09}.
The third state is extremely weakly bound and so may be an artefact of our
truncated action.

In the unitary limit, the evolution generates an infinite number of four-body 
resonant states during each Efimov cycle, although it seems unlikely that all of 
these persist to the physical limit. These states 
lie just below each atom-trimer threshold and follow a double-logarithmic, 
or ``super-Efimov" pattern \cite{mns13}. They obey a universal scaling relation 
analogous to that of Ref.~\cite{had11}. However the scales at which they appear 
are independent of the initial conditions on the four-body couplings. This 
supports the conclusion of Refs.~\cite{phm04,hp07} that there is no additional 
relevant parameter in four-boson systems with contact interactions.

\vspace{10pt}
\begin{acknowledgments}
We are grateful to S. Floerchinger, B. Krippa, S. Moroz, R. Schmidt and N. Walet for 
helpful discussions. BJA acknowledges support from CONACyT.
\end{acknowledgments}

\end{document}